\documentclass[11pt]{article}

\newif\ifsubmission
\submissionfalse       % <-- use this line for working-paper/public version

\usepackage[authoryear]{natbib}
\usepackage{amsmath,amssymb,amsthm}
\usepackage{dsfont}
\usepackage{hyperref}
\usepackage{graphicx}
\usepackage{enumitem}
\usepackage{booktabs}
\ifsubmission
  \usepackage{lineno}
  \modulolinenumbers[5]
\else
\fi

\newcommand{\Prb}{\mathbb{P}}

\newcommand{\High}{\mathrm{H}}
\newcommand{\Low}{\mathrm{L}}
\newcommand{\Easy}{\mathrm{E}}
\newcommand{\Hard}{\mathrm{D}}
\newcommand{\dd}{\,\mathrm{d}}

\newtheorem{assumption}{Assumption}

\newtheorem{theorem}{Theorem}
\newtheorem{lemma}{Lemma}

\newtheorem{proposition}{Proposition}
\newtheorem{corollary}{Corollary}
\newtheorem{example}{Example}
\newtheorem{remark}{Remark}

\begin{document}

\title{Diagnostic Feedback under Hidden Task Difficulty\thanks{We are grateful to Emin Ablyatifov, Jacob Gershman, Arthur Izgarshev, David Li, Anastasia Makhmudova, and Yuliia Tukmakova for helpful comments and suggestions. All remaining errors are our own.}}
\author{Mark Izgarshev\\ \small School number 1239, Vspolny per., 6, Moscow, Russia
\and
Georgy Lukyanov\\ \small Toulouse School of Economics, 1, Esplanade de l'Universit\'{e}, Toulouse, France}
\date{}
\maketitle

\begin{abstract}
An assessment is meant to tell an agent about herself, but its purchase can reveal the task. An evaluator privately observes task difficulty and may publicly purchase an ability diagnostic before effort. With crossing returns across task--ability pairs, adoption and the result jointly shape effort. When high ability is common, difficult tasks are rare, and diagnostic cost is intermediate, an equilibrium refinement uniquely selects diagnosis on the difficult task. Neither type diagnoses when difficulty is public or adoption cannot depend on it. Thus hidden difficulty can produce information about the agent. The mechanism survives continuous effort and imperfect diagnostics.
\end{abstract}



\section{Introduction}
\label{sec:introduction}

An agent about to prepare for something is often assessed first: a diagnostic exam before a demanding course, a skills audit before a project is staffed, a screening exercise before a trainee is handed an assignment. Each of these instruments is built to tell the agent about herself. But somebody had to decide to run it, and that somebody usually knows more about the assignment than she does. The decision to assess is then a second signal, carried by the same instrument, and it points at the task rather than at the person.

That second signal has content because a diagnostic is not equally useful everywhere. On a demanding task it can identify the agents for whom extra preparation will actually change the outcome. On an undemanding task the same instrument may not be worth what it costs, precisely because it will change nobody's behavior. Adoption can then reveal that the task is demanding---and the agent, who reads her result in the light of that inference, responds to the two signals jointly rather than to either one alone.

The question we ask is whether this second signal can make an assessment worth buying that would not otherwise be bought, and what it does to the way the agent reads her own result.\footnote{We take the diagnostic technology as given and ask when it is used, rather than asking how it should be designed. That division of labor is deliberate: the object of interest here is the adoption margin, and holding the instrument fixed is what isolates it.}

We work with two tasks and two ability levels. The evaluator privately observes whether the task is easy or difficult; neither player initially knows the agent's ability. Before the agent chooses costly effort, the evaluator may publicly purchase a diagnostic that reveals ability. The marginal return to effort crosses across task--ability pairs: effort is most productive for low ability on an easy task and for high ability on a difficult one. This is not an assumption about who is more productive overall, and Section~\ref{subsec:foundation} shows how a standard threshold success technology generates it: effort matters most for the agent who sits nearest the requirement.

Suppose high ability is common, difficult tasks are rare, and the diagnostic cost lies in an intermediate range. We first show that there is a separating equilibrium in which only the difficult-task evaluator diagnoses: adoption reveals the difficult task, the high-ability agent then works, and the low-ability agent does not. Pooling on no diagnosis can be sustained as well, but only by an off-path belief that attributes an unexplained diagnostic to the type who could never want one. D1 removes that belief, and separation is then the unique equilibrium---not merely the unique pure-strategy one, since neither evaluator type can be made indifferent between diagnosing and not.

We show, second, that this activity is not a value-of-information effect in disguise. If task difficulty is public, neither evaluator diagnoses. On the easy task the diagnostic motivates too few agents to pay for itself; on the difficult task unresolved ability is an asset, because without a diagnostic everyone works, whereas screening stops the low-ability agents. The same evaluator also declines to diagnose if she must fix her policy before observing difficulty. Positive adoption under hidden difficulty, set against these two benchmarks, is what we call \emph{diagnostic over-adoption}: a comparison across institutions, and not yet a claim about welfare.

Common sense would suggest two things here, and neither survives. The first is that concealing one thing should not manufacture information about another---that opacity about the task ought to reduce, or at worst leave alone, the information produced about the agent. It does the opposite: the diagnostic is purchased with positive probability exactly when difficulty is hidden, and with probability zero when it is public. The second is that favorable feedback motivates and unfavorable feedback discourages. The sign is not a property of the label. On an easy task it is the low-ability agent for whom effort is pivotal, so an unfavorable result is the one that induces work. Which reading applies is settled by what adoption has revealed, which is why the two signals cannot be separated.

A third guess is worth disposing of immediately, since it is the natural objection to the whole exercise: if the evaluator wants effort on the difficult task, why not simply say so? Because she cannot be believed. An announcement that induces effort is an announcement the easy-task evaluator wants to imitate, and imitation is free. Buying a diagnostic is not free, and the ability result it produces then restricts who responds. A costly action that generates an outcome can separate types where a claim about the state cannot.

The paper is related first to models in which an informed principal's intervention or feedback changes effort by revealing private information \citep{BenabouTirole2002,BenabouTirole2003,Rosaz2012,LilgeRamchandani2024}. The closest conceptual precedent is \citet{BenabouTirole2003}, where an intervention may reveal the principal's confidence in the agent or her information about the task. We separate the decision to create self-knowledge from the content of the resulting feedback. That separation is what produces an adoption signal, reverses the motivating result across tasks, and opens a wedge between hidden difficulty, public difficulty, and commitment.

The paper also speaks to work on exams, grades, and performance feedback. \citet{Shimoji2023} treats exam setting as an information-design problem, while \citet{Ederer2010}, \citet{ErtacEtAl2016}, and \citet{FischerSliwka2018} study the incentive effects of feedback. There the policy instrument is normally the content, precision, or timing of what is disclosed. Here the technology is fixed; what is endogenous and observable is whether it is used at all, and the evaluator's private information is about the task rather than about the agent.\footnote{The distinction matters for interpretation. In much of the feedback literature the principal knows something about the agent that the agent does not. Here she knows something about the environment, and the agent's own characteristic is unknown to both---which is why a diagnostic can be informative to the principal's audience and to the principal at the same time.}

Finally, our evaluator is an informed designer without ex ante commitment, which connects the paper to informed persuasion and endogenous information structures \citep{Hedlund2017,Hedlund2024,KoesslerSkreta2023,Zapechelnyuk2023}, and to testing and evidence \citep{Herresthal2022,FigueroaGuadalupi2023}. The nearest information-acquisition papers differ in what is learned and in whether the act of learning is observed. In \citet{Kaya2010} a principal may postpone learning a productivity parameter because being informed carries a signaling cost in a continuing relationship. In \citet{EkmekciKos2023} a sender covertly learns his own type before taking a conventional signaling action. Our evaluator already knows her type---the task---and observably acquires information about the \emph{receiver}. Acquisition signals the task; the information acquired changes what the receiver does. \citet{Hedlund2024} studies signaling through the choice of a persuasion experiment, and the public test in \citet{FigueroaGuadalupi2023} concerns the informed sender's own product quality. Here the object tested is a separate characteristic of the receiver whose relevance is determined by the sender's hidden task. The present paper tries to fill in that gap.

What makes the mechanism work is that the diagnostic never directly discloses what the evaluator knows. Its adoption reveals the task in equilibrium; its result gives the agent self-knowledge. Keeping both dimensions binary buys a complete characterization of the equilibrium set and a transparent application of D1 \citep{BanksSobel1987,ChoKreps1987}, and Sections~\ref{sec:continuous}--\ref{sec:noisy} show that the adoption signal and the comparison with transparency both survive continuous effort and diagnostic error.

\medskip
Section~\ref{sec:model} presents the environment and its threshold foundation. Section~\ref{sec:feedback} derives the feedback reversal and explains why cheap talk cannot substitute for diagnosis. Section~\ref{sec:hidden} characterizes equilibrium under hidden difficulty and applies D1. Section~\ref{sec:benchmarks} develops the public-difficulty and commitment benchmarks. Section~\ref{sec:continuous} allows continuous effort, and Section~\ref{sec:noisy} allows diagnostic error. Sections~\ref{sec:symmetry}--\ref{sec:welfare} treat symmetry, welfare, and empirical content. Section~\ref{sec:conclusion} concludes. Proofs are in Appendix~\ref{app:proofs}.

\section{Environment}
\label{sec:model}

The model has to do two things at once: let the diagnostic be informative about the agent, and let the decision to buy it be informative about the task. The first requires ability to be unknown to both players; the second requires the evaluator to know something the agent does not. Everything else is kept as small as possible.

\subsection{Tasks, ability, and effort}

There are two players: an evaluator \(P\) and an agent \(A\). Nature draws a task
\[
 t\in\{\Easy,\Hard\},
 \qquad
 \Prb(t=\Hard)=\rho\in(0,1),
\]
where \(\Easy\) denotes an easy task and \(\Hard\) a difficult one. The evaluator observes \(t\); the agent does not. Nature independently draws the agent's ability
\[
 \theta\in\{\Low,\High\},
 \qquad
 \Prb(\theta=\High)=\pi\in(1/2,1).
\]
Initially neither player observes \(\theta\).\footnote{Ability here is task-relevant readiness or fit, not necessarily a stable trait the agent already knows. The diagnostic matters because this component of productivity is latent for both sides; if the agent knew it already, the instrument would no longer create self-knowledge and the economic question would change.}

The agent chooses binary effort \(e\in\{0,1\}\). Effort costs \(c>0\) and raises the probability of success by \(\Delta_{t\theta}\), according to the crossing specification
\begin{equation}
\label{eq:increments}
\begin{array}{c@{\qquad}cc}
 & \theta=\Low & \theta=\High\\
\toprule
t=\Easy & a & b\\
t=\Hard & b & a\\
\bottomrule
\end{array}
\qquad\text{with}\qquad
0<b<a.
\end{equation}
Let \(p_{t\theta}\) be the success probability at \(e=0\), and suppose
\[
0\leq p_{t\theta}\leq 1-\Delta_{t\theta},
\]
so that success occurs with probability \(p_{t\theta}+e\Delta_{t\theta}\). The level terms \(p_{t\theta}\) are free to let high ability raise success on either task; only the increments govern the effort and diagnostic decisions.

The agent values success at \(v>0\), and the evaluator's value of success is normalized to one. Expected payoffs are
\begin{equation}
\label{eq:payoffs}
u_A=v\Prb(\text{success})-ce,
\qquad
u_P=\Prb(\text{success})-Kz,
\end{equation}
where \(z\in\{0,1\}\) is the diagnostic decision and \(K>0\) its cost. Write
\[
x\equiv \frac{c}{v}
\]
for the agent's effort threshold: she works whenever the expected incremental success probability exceeds \(x\).\footnote{The evaluator does not internalize the agent's effort cost, so the two players disagree about how much effort is desirable. We return to this in Section~\ref{sec:welfare}, where surplus is computed with \(c\) and \(K\) both treated as real resource costs.}

\subsection{The diagnostic and timing}

After observing \(t\), the evaluator chooses either no diagnostic, \(z=0\), or a diagnostic, \(z=1\). The decision is public. In the baseline model the diagnostic perfectly reveals \(\theta\) to both players and has no direct effect on success. The timing is:

\begin{enumerate}[label=(\arabic*)]
 \item Nature draws \((t,\theta)\); the evaluator observes \(t\).
 \item The evaluator publicly chooses \(z\in\{0,1\}\).
 \item If \(z=1\), the diagnostic publicly reveals \(\theta\).
 \item The agent chooses \(e\in\{0,1\}\), and success is realized.
\end{enumerate}

The evaluator cannot commit ex ante to a task-contingent diagnostic policy; Section~\ref{sec:benchmarks} studies public difficulty and state-independent commitment in turn.\footnote{The diagnostic is purely informational: a bad result carries no penalty of its own, and nothing about the agent's access to the task depends on it. This is the case in which the adoption margin is cleanest, since any direct consequence of the result would give the evaluator a second, non-informational reason to buy. It also fits the leading applications---a practice exam, a preparatory audit---better than it fits a licensing threshold, where failing is itself the sanction.}

Let
\[
\mu_z\equiv \Prb(t=\Hard\mid z)
\]
be the agent's posterior about task difficulty after the diagnostic decision. Because ability and difficulty are independent, the diagnostic result does not move \(\mu_z\). If no diagnostic is used, the expected marginal product of effort is
\begin{align}
m_{\Easy}&\equiv (1-\pi)a+\pi b, \label{eq:mE}\\
m_{\Hard}&\equiv (1-\pi)b+\pi a, \label{eq:mH}\\
D_0(\mu)&\equiv (1-\mu)m_{\Easy}+\mu m_{\Hard}. \label{eq:D0}
\end{align}
If ability is revealed, the corresponding marginal products are
\begin{align}
D_{\Low}(\mu)&\equiv (1-\mu)a+\mu b
 =a-\mu(a-b), \label{eq:DL}\\
D_{\High}(\mu)&\equiv (1-\mu)b+\mu a
 =b+\mu(a-b). \label{eq:DH}
\end{align}
So after no diagnostic the agent works if \(D_0(\mu_0)>x\), and after a diagnostic an agent revealed to have ability \(\theta\) works if \(D_\theta(\mu_1)>x\). Note that \(D_{\Low}\) falls and \(D_{\High}\) rises in \(\mu\), which is the crossing of \eqref{eq:increments} restated in beliefs: news about the task moves the two ability types in opposite directions. Every parameter restriction below is strict, so tie-breaking is immaterial.

\subsection{Parameter region}

The results are stated for a population in which high ability is common and difficult tasks are rare. Section~\ref{sec:symmetry} shows what happens when the modal ability type is reversed, and the reversal is exact.

\begin{assumption}[Crossing incentives]
\label{ass:crossing}
\[
0<b<x<a,
\qquad
m_{\Easy}<x<m_{\Hard}.
\]
\end{assumption}

The first pair of inequalities says that when both difficulty and ability are known, effort is worth supplying exactly for the matched pairs \((\Easy,\Low)\) and \((\Hard,\High)\). The second pair says that without ability information effort is unattractive on the easy task and attractive on the difficult one.

\begin{assumption}[Rare difficult tasks]
\label{ass:rare}
\[
D_0(\rho)<x,
\qquad
D_{\High}(\rho)<x<D_{\Low}(\rho).
\]
\end{assumption}

Before the diagnostic decision conveys anything, then, the agent does not work while ability is unresolved; and if a diagnostic is used in a way that carries no news about the task, only a low-ability agent works. These inequalities hold whenever
\begin{equation}
\label{eq:rho-bound}
\rho<
\bar\rho\equiv
\min\left\{
\frac{x-m_{\Easy}}{m_{\Hard}-m_{\Easy}},
\frac{x-b}{a-b},
\frac{a-x}{a-b}
\right\},
\end{equation}
and Assumption~\ref{ass:crossing} makes every term in \(\bar\rho\) positive, so the region is not empty.

\begin{assumption}[Intermediate diagnostic cost]
\label{ass:cost}
\[
m_{\Easy}<K<\pi a.
\]
\end{assumption}

The lower bound is deliberately stronger than the easy evaluator's on-path incentive constraint, which would only require \(K>\pi b\). It says that even if a diagnostic somehow induced effort from every ability type on the easy task, the resulting gain would not cover the cost.\footnote{Assuming the stronger bound is what makes the refinement in Section~\ref{sec:hidden} robust rather than delicate: it is a statement about the easy type's payoff under \emph{every} continuation, not only under the one she faces in equilibrium. The price is that the separating region is described by a sufficient condition rather than a necessary one.} The upper bound says that a difficult-task evaluator is willing to diagnose when doing so induces only the high-ability agents to work.

\subsection{A threshold foundation}
\label{subsec:foundation}

Read carelessly, \eqref{eq:increments} looks like a claim that low ability is better on easy tasks. It is not: it is a statement about the \emph{marginal} return to effort, and it is what a standard threshold technology produces.

Suppose success has probability
\begin{equation}
\label{eq:threshold-tech}
F(\alpha_\theta+\gamma e-T_t),
\end{equation}
where \(\alpha_{\High}>\alpha_{\Low}\), \(\gamma>0\), and \(F\) has a continuous density \(f\) that is symmetric and strictly unimodal around zero. The marginal effect of effort at index \(q=\alpha_\theta-T_t\) is
\[
\Delta(q)=F(q+\gamma)-F(q)
 =\int_q^{q+\gamma}f(s)\dd s.
\]

\begin{lemma}
\label{lem:foundation}
Let
\[
T_{\Easy}=\alpha_{\Low}+\frac{\gamma}{2},
\qquad
T_{\Hard}=\alpha_{\High}+\frac{\gamma}{2}.
\]
Then the incremental success probabilities generated by \eqref{eq:threshold-tech} satisfy \eqref{eq:increments}, with
\[
a=\Delta\!\left(-\frac{\gamma}{2}\right)
\quad\text{and}\quad
b=\Delta\!\left(\alpha_{\High}-\alpha_{\Low}-\frac{\gamma}{2}\right)
 =\Delta\!\left(\alpha_{\Low}-\alpha_{\High}-\frac{\gamma}{2}\right)
 <a.
\]
\end{lemma}

The mechanics are worth stating in words. The easy task puts low ability nearest the success requirement; the difficult task does the same for high ability. Since extra effort shifts the success index by a fixed amount, it does the most good for whichever ability type happens to sit closest to the requirement, and the crossing follows. The same primitive also lets high ability raise the \emph{level} of success on both tasks even where its marginal return to effort is the lower of the two---so nothing here says that being able is a disadvantage.

Lemma~\ref{lem:foundation} produces a matrix in which the two matched cells share a value and the two mismatched cells share another. That coincidence is an artifact of the thresholds chosen; the main mechanism does not depend on it.

\begin{remark}
\label{rem:asymmetric}
Replace \eqref{eq:increments} with four distinct increments
\[
\Delta_{\Easy\Low}=a_{\Easy},
\quad
\Delta_{\Hard\High}=a_{\Hard},
\quad
\Delta_{\Easy\High}=b_{\Easy},
\quad
\Delta_{\Hard\Low}=b_{\Hard},
\]
and impose the task-specific counterparts of the maintained assumptions:
\[
\max\{b_{\Easy},b_{\Hard}\}<x<\min\{a_{\Easy},a_{\Hard}\},
\qquad
m_{\Easy}<x<m_{\Hard},
\]
where \(m_{\Easy}=(1-\pi)a_{\Easy}+\pi b_{\Easy}\) and \(m_{\Hard}=(1-\pi)b_{\Hard}+\pi a_{\Hard}\). Retain Assumption~\ref{ass:rare} after replacing
\[
D_{\Low}(\mu)=(1-\mu)a_{\Easy}+\mu b_{\Hard},
\qquad
D_{\High}(\mu)=(1-\mu)b_{\Easy}+\mu a_{\Hard},
\]
and replace Assumption~\ref{ass:cost} by \(m_{\Easy}<K<\pi a_{\Hard}\). The equilibrium, refinement, benchmark, noisy-diagnostic, and welfare arguments then go through with the corresponding task-specific substitutions. The nonempty-cost condition \eqref{eq:pi-bound} becomes
\[
\pi>\frac{a_{\Easy}}{a_{\Easy}+a_{\Hard}-b_{\Easy}},
\]
and \eqref{eq:welfare-diff} becomes \((1-\pi)\big[(1+v)b_{\Hard}-c\big]+K\). In the continuous-effort extension, the largest effort any continuation can elicit is \(\eta(\max\{a_{\Easy},a_{\Hard}\})\), and the first lower bound in \eqref{eq:continuous-K} must use that quantity. We retain the two-parameter form because it halves the notation and because Lemma~\ref{lem:foundation} delivers it exactly.
\end{remark}

\section{Feedback and motivation}
\label{sec:feedback}

Before turning to the adoption decision it is worth establishing what a diagnostic result does once the task is known, because the answer is not the one the words ``favorable'' and ``unfavorable'' suggest.

\subsection{A feedback reversal}

Call the result \(\theta=\High\) favorable feedback and \(\theta=\Low\) unfavorable feedback. The crossing technology makes the motivational force of these labels depend on the task.

\begin{proposition}
\label{prop:reversal}
Under Assumption~\ref{ass:crossing}, if task difficulty is known, then:
\begin{enumerate}[label=(\roman*)]
 \item on the easy task, a low-ability agent works after a diagnostic and a high-ability agent does not;
 \item on the difficult task, a high-ability agent works after a diagnostic and a low-ability agent does not.
\end{enumerate}
Without a diagnostic, the agent does not work on the easy task and does work on the difficult one.
\end{proposition}

The proposition separates what ability does to the level of success from what it does to the return on effort. On an easy task, high ability can raise expected success while lowering the value of working harder, because the agent is already comfortably above the requirement; a low-ability agent is still close enough for effort to be pivotal. On a difficult task the positions are exchanged.

Feedback labels therefore carry no context-free motivational meaning. A favorable result induces effort when adoption reveals a difficult task, and fails to induce it when adoption reveals an easy one---the same message, read against two different inferences. Field and laboratory studies of feedback often find heterogeneous, sometimes sign-reversing effects on subsequent effort \citep{ErtacEtAl2016,FischerSliwka2018}. Here the heterogeneity is not a nuisance parameter but the prediction, and it is indexed by something observable: the difficulty of the task for which the feedback was provided.

\subsection{Why an announcement is not enough}

The obvious objection is that the evaluator could simply announce the difficulty. Suppose then, as a benchmark, that the diagnostic is unavailable and she may instead send a costless, unverifiable message before effort.

\begin{proposition}
\label{prop:cheap-talk}
Under Assumption~\ref{ass:crossing} there is no fully revealing cheap-talk equilibrium. Under Assumption~\ref{ass:rare} a babbling equilibrium induces no effort.
\end{proposition}

The reason is that both types want the same thing from the agent. If a message credibly conveyed a difficult task the agent would work, since \(m_{\Hard}>x\); and the easy evaluator would then strictly prefer to send it, because effort raises her success probability by \(m_{\Easy}>0\) at no cost. Diagnostic adoption is different on both counts. Imitating it requires paying \(K\), and the realized ability result then restricts who actually works. It is the conjunction---costly to imitate, and consequential for the continuation---that lets the action separate types where a bare claim about the state cannot.

\section{Hidden task difficulty}
\label{sec:hidden}

We can now put the two pieces together. The evaluator's decision is a signal about the task; the diagnostic result is a signal about the agent; and the agent's response depends on both.

\subsection{A separating equilibrium}

Consider the assessment policy
\begin{equation}
\label{eq:sep-policy}
z(\Easy)=0,
\qquad
z(\Hard)=1.
\end{equation}
On path, no diagnostic reveals an easy task and a diagnostic reveals a difficult one. The agent therefore does not work after \(z=0\), while after \(z=1\) a favorable result induces effort and an unfavorable one does not.

Since baseline success does not depend on \(z\), it is enough to compare the success gain created by effort, net of \(K\). Under \eqref{eq:sep-policy} the difficult evaluator obtains
\[
\pi a-K
\]
relative to choosing no diagnostic and inducing no effort, while an easy evaluator who imitates her obtains
\[
\pi b-K.
\]
Assumption~\ref{ass:cost} makes the first positive and the second negative, the latter because \(K>m_{\Easy}>\pi b\).

\begin{proposition}
\label{prop:separating}
Under Assumptions~\ref{ass:crossing} and~\ref{ass:cost} there is a Perfect Bayesian equilibrium in which only the difficult-task evaluator purchases the diagnostic. Beliefs are
\[
\mu_0=0,
\qquad
\mu_1=1,
\]
and effort is
\[
e(0)=0,
\qquad
e(1,\High)=1,
\qquad
e(1,\Low)=0.
\]
\end{proposition}

Adoption and the result do different jobs, and it takes both. Adoption identifies the task; conditional on that inference, the result identifies whether effort is productive. Neither signal alone implements the allocation: cheap talk cannot reveal the task credibly, and a diagnostic that carries no news about the task is read under the prior and motivates the wrong ability type throughout the region of Assumption~\ref{ass:rare}.

\subsection{The equilibrium set}

Separation is not the only equilibrium. If both types choose no diagnostic, Bayes' rule gives \(\mu_0=\rho\), so the agent does not work by Assumption~\ref{ass:rare}. Pooling can then be sustained by attributing an unexpected diagnostic to the easy type, \(\mu_1=0\), in which case only an unfavorable result induces effort and the two deviation gains are
\[
(1-\pi)a-K<0
\qquad\text{and}\qquad
(1-\pi)b-K<0,
\]
both negative because \(K>m_{\Easy}>(1-\pi)a>(1-\pi)b\).

\begin{proposition}
\label{prop:pbe-set}
Under Assumptions~\ref{ass:crossing}--\ref{ass:cost}:
\begin{enumerate}[label=(\roman*)]
 \item the separating equilibrium of Proposition~\ref{prop:separating} exists;
 \item a pooling equilibrium in which both types choose no diagnostic exists for suitable off-path beliefs;
 \item there is no pure-strategy equilibrium in which both types diagnose, nor one in which only the easy-task evaluator diagnoses.
\end{enumerate}
\end{proposition}

Both of the profiles ruled out in part~(iii) fail for the same reason. If adoption does not identify the difficult task, then under the resulting belief only an unfavorable result motivates effort; the easy evaluator's gain is then at most \((1-\pi)a-K<0\), and she prefers to save the cost. The bound is generous to her---it credits her with the most favorable continuation available---which is why the argument leaves no room for the reverse separating profile either.

It is natural to ask whether randomization opens up anything further, particularly since mixed strategies matter in many signaling games. Here it does not, and the reason is instructive.

\begin{proposition}
\label{prop:no-mixing}
Under Assumptions~\ref{ass:crossing}--\ref{ass:cost}, no Perfect Bayesian equilibrium has either evaluator type randomize over the diagnostic decision. The set of Perfect Bayesian equilibria therefore consists of the separating profile of Proposition~\ref{prop:separating} together with pooling on no diagnosis.
\end{proposition}

For the easy type the argument does not depend on beliefs at all: no diagnostic strictly dominates diagnosing, against every continuation the agent could conceivably play, because the largest gain a diagnostic could deliver her is \(m_{\Easy}<K\). For the difficult type indifference is also unavailable, but for a different reason. If she randomized, then no diagnostic would be observed with positive probability and would carry a posterior no higher than the prior, so it would induce no effort at all; diagnosing, meanwhile, would be attributed to her with certainty and pay \(\pi a-K>0\). There is no interior probability at which those two are equal.

\subsection{D1 selection}

Pooling survives only on the strength of an off-path belief that attributes an unexplained diagnostic to the type who could never want one. That is exactly what D1 forbids.

For any continuation response after \(z=1\)---including any randomization by the agent---the easy evaluator's success gain is a weighted average of the increments in the easy row of \eqref{eq:increments}, and is therefore largest when both ability types work. Assumption~\ref{ass:cost} then gives
\begin{equation}
\label{eq:easy-never}
\sup_{\text{continuation responses}}
\left\{\text{easy evaluator's gain from }z=1\right\}
\leq m_{\Easy}-K<0.
\end{equation}
The easy evaluator cannot profit from an unexpected diagnostic under any response whatsoever. The difficult evaluator can, as soon as the deviation is attributed to her: adoption then implies \(\mu_1=1\), only high ability works, and the gain is \(\pi a-K>0\).

D1 assigns zero posterior probability to a sender type whose set of profitable deviation responses is strictly smaller in this sense \citep{BanksSobel1987}. Following an unexpected diagnostic at the pooling profile, the agent must therefore place probability one on a difficult task---and that response is precisely what makes the deviation worth taking.

\begin{theorem}
\label{thm:d1}
Under Assumptions~\ref{ass:crossing}--\ref{ass:cost}, the separating equilibrium of Proposition~\ref{prop:separating} is the unique Perfect Bayesian equilibrium satisfying D1.
\end{theorem}

Two features of this are worth flagging, because refinement arguments are often fragile. First, nothing turns on a delicate ranking of marginal deviations: the lower cost bound rules out a profitable diagnostic for the easy type under \emph{every} continuation response, and the upper bound leaves one available to the difficult type. Both comparisons are strict, which is why they will survive small diagnostic errors in Section~\ref{sec:noisy}. Second, the uniqueness claim is over all equilibria and not only the pure ones, by Proposition~\ref{prop:no-mixing}---so the refinement is doing the whole of the selection work, with no auxiliary restriction on strategies standing behind it.\footnote{In the baseline model, the evaluator has two types and finitely many actions, and the agent has finitely many continuation responses. The agent's independently drawn ability enters as a lottery over continuation payoffs rather than as a second evaluator type.}

\subsection{Existence and comparative statics}

The restrictions are jointly feasible, and reading off how the region moves is straightforward.

\begin{proposition}
\label{prop:region}
Fix \(0<b<x<a\) and \(\pi>1/2\).
\begin{enumerate}[label=(\roman*)]
 \item The cost interval in Assumption~\ref{ass:cost} is nonempty if and only if
 \begin{equation}
 \label{eq:pi-bound}
 \pi>\frac{a}{2a-b}.
 \end{equation}
 \item Conditional on \eqref{eq:pi-bound}, that interval expands strictly with \(\pi\): its lower endpoint \(m_{\Easy}=a-\pi(a-b)\) falls and its upper endpoint \(\pi a\) rises.
 \item If Assumption~\ref{ass:crossing} holds, then Assumption~\ref{ass:rare} holds for every \(\rho\in(0,\bar\rho)\), with \(\bar\rho\) as in \eqref{eq:rho-bound}.
\end{enumerate}
\end{proposition}

The modal ability type is what determines which task can finance a diagnostic. A larger \(\pi\) works from both ends: it increases the mass of high-ability agents who respond to favorable feedback on the difficult task, and it lowers the expected marginal return to effort on the easy one. Rarity of difficult tasks does the complementary job, ensuring that a diagnostic carrying no news about the task is read mainly through the easy-task environment.

\begin{example}
\label{ex:numeric}
Let
\[
a=0.8,\quad b=0.2,\quad x=0.5,\quad
\pi=0.9,\quad \rho=0.2,\quad K=0.4.
\]
Then
\[
m_{\Easy}=0.26,\qquad
m_{\Hard}=0.74,\qquad
D_0(\rho)=0.356,
\]
and
\[
D_{\High}(\rho)=0.32<0.5<0.68=D_{\Low}(\rho).
\]
The difficult evaluator's gain from diagnosis in the separating equilibrium is
\[
\pi a-K=0.32,
\]
against \(\pi b-K=-0.22\) for an easy evaluator who imitates her. Here \(\bar\rho=1/2\) and \(a/(2a-b)=4/7\), so \eqref{eq:rho-bound} and \eqref{eq:pi-bound} hold with room to spare, and all assumptions are strict.
\end{example}

\section{Transparency and commitment benchmarks}
\label{sec:benchmarks}

If adoption is doing the work we claim, then removing its informational content should remove the adoption. This section removes it in two different ways.

\subsection{Public task difficulty}

Suppose first that \(t\) is public before the evaluator chooses \(z\). On the easy task, no diagnostic produces no effort because \(m_{\Easy}<x\), while a diagnostic induces only low ability to work and yields
\[
(1-\pi)a-K<0
\]
relative to no effort. The easy evaluator therefore does not diagnose.

On the difficult task, no diagnostic induces everyone to work because \(m_{\Hard}>x\), for an incremental success gain of \(m_{\Hard}\). A diagnostic induces only high ability to work and yields \(\pi a-K\). The difference in favor of doing nothing is
\begin{equation}
\label{eq:public-hard-diff}
m_{\Hard}-(\pi a-K)=(1-\pi)b+K>0,
\end{equation}
and its two terms name the two costs of diagnosing: the low-ability effort it destroys, and the fee.

\begin{proposition}
\label{prop:public}
Under Assumptions~\ref{ass:crossing} and~\ref{ass:cost}, if task difficulty is public then neither evaluator type purchases the diagnostic.
\end{proposition}

Transparency thus eliminates diagnosis on both tasks, but for opposite reasons. On the easy task the diagnostic motivates too few agents to pay for itself. On the difficult task it reveals too much: ability uncertainty was doing useful work, and resolving it stops the low-ability agents who would otherwise have supplied effort.\footnote{The second reason is the more interesting one, because it is a case in which an evaluator strictly prefers her audience to remain uninformed about something she could cheaply inform them about. It is also what makes the comparison in Corollary~\ref{cor:overadoption} a statement about institutions rather than about the value of information: the same instrument at the same price is bought in one regime and not the other.}

\subsection{State-independent commitment}

Now suppose that, before observing difficulty, the evaluator can commit to a diagnostic policy that cannot depend on \(t\)---equivalently, imagine a central rule requiring the same assessment decision for both tasks. Committing to no diagnostic leaves the agent with the prior \(\rho\) and no effort, by Assumption~\ref{ass:rare}.

Committing to diagnose conveys nothing about difficulty, so an unfavorable result induces effort and a favorable one does not. The ex ante success gain is therefore
\begin{equation}
\label{eq:commit-gain}
(1-\pi)\big[(1-\rho)a+\rho b\big]-K,
\end{equation}
which is negative because
\[
(1-\pi)\big[(1-\rho)a+\rho b\big]
<(1-\pi)a<m_{\Easy}<K.
\]

\begin{proposition}
\label{prop:commitment}
Under Assumptions~\ref{ass:crossing}--\ref{ass:cost}, an evaluator restricted to a state-independent diagnostic policy commits to no diagnostic. The same holds if she may randomize with a probability that does not depend on task difficulty.
\end{proposition}

Randomizing cannot help, because a state-independent diagnostic remains uninformative about \(t\) however often it is used, and the payoff is linear in the probability of using it. What the commitment benchmark isolates is that the value of the instrument here is not in its output but in the inference its purchase supports.

\begin{corollary}
\label{cor:overadoption}
In the unique D1 equilibrium under hidden difficulty, the diagnostic is purchased with ex ante probability \(\rho>0\). It is never purchased under public difficulty or under state-independent commitment.
\end{corollary}

Hidden information about the task therefore creates diagnostic activity rather than suppressing it. The difficult evaluator buys the technology because buying it separates her from the easy evaluator; strip out that inference, in either of the two ways above, and the technology is not worth its price.

\section{Continuous effort}
\label{sec:continuous}

Binary effort makes the extensive margin transparent, which is what the argument so far has turned on. It also leaves open whether the conclusion is an artifact of an all-or-nothing response, so we now let effort vary continuously with perceived productivity.

Let \(e\in[0,\bar e]\) with \(\bar e\leq1\), and let success occur with probability
\begin{equation}
\label{eq:continuous-success}
p_{t\theta}+e\Delta_{t\theta},
\end{equation}
with \(\Delta\) as in \eqref{eq:increments} and \(\bar e\) chosen so that \eqref{eq:continuous-success} remains a probability. The agent's effort cost is
\begin{equation}
\label{eq:continuous-cost}
C(e)=vxe+\frac{\kappa}{2}e^2,
\qquad \kappa>0,
\end{equation}
and we assume
\[
\bar e>\frac{v(a-x)}{\kappa},
\]
so the upper bound never binds. If the agent perceives the marginal success productivity of effort to be \(q\), her unique optimal effort is
\begin{equation}
\label{eq:continuous-br}
\eta(q)
\equiv
\frac{v}{\kappa}(q-x)_+.
\end{equation}
The same threshold \(x\) still decides whether effort is zero; above it, effort now varies with perceived productivity.

For \(e\in[0,1]\), \eqref{eq:continuous-success} is a convex combination of success without and with the one-unit intervention of Section~\ref{subsec:foundation}, so continuous effort can be read as the intensity, or implementation probability, of that intervention. The productivity matrix keeps the same threshold foundation.

Retain Assumptions~\ref{ass:crossing} and~\ref{ass:rare}, and write
\[
\eta_a\equiv\eta(a),
\qquad
\eta_{\Hard}\equiv\eta(m_{\Hard}).
\]
Replace Assumption~\ref{ass:cost} with
\begin{equation}
\label{eq:continuous-K}
\max\left\{
m_{\Easy}\eta_a,\,
\pi a\eta_a-m_{\Hard}\eta_{\Hard}
\right\}
<K<\pi a\eta_a.
\end{equation}
The three bounds have the three jobs one would expect: the first lower bound makes diagnosis unprofitable for the easy evaluator under every sequentially rational continuation, the second makes no diagnosis preferable on a public difficult task, and the upper bound makes diagnosis profitable for the difficult evaluator when adoption reveals difficulty.

\begin{theorem}
\label{thm:continuous}
Suppose Assumptions~\ref{ass:crossing} and~\ref{ass:rare} and condition \eqref{eq:continuous-K} hold in the continuous-effort model.
\begin{enumerate}[label=(\roman*)]
 \item The unique D1 equilibrium has the easy evaluator choose no diagnostic and the difficult evaluator diagnose. After diagnosis, high ability supplies effort \(\eta_a\) and low ability supplies zero.
 \item If task difficulty is public, neither evaluator diagnoses. Effort is zero on the easy task and \(\eta_{\Hard}\) on the difficult one.
 \item Under state-independent commitment the evaluator chooses no diagnostic.
\end{enumerate}
The interval in \eqref{eq:continuous-K} is nonempty whenever \(m_{\Easy}<\pi a\), equivalently whenever \eqref{eq:pi-bound} holds.
\end{theorem}

Continuous effort pulls apart two effects that the binary model runs together, and the second lower bound in \eqref{eq:continuous-K} is where they meet. Under public difficulty, unresolved ability elicits effort \(\eta_{\Hard}\) from everyone; diagnosing raises the high-ability agents to \(\eta_a\) but zeroes out the low-ability agents and costs \(K\). Whether that trade is worth making is now a quantitative question rather than a qualitative one, and the bound is what answers it. Under hidden difficulty no diagnosis is instead read as an easy task and elicits nothing, so the difficult evaluator diagnoses in order to separate.

\begin{example}
\label{ex:continuous}
Retain
\[
a=0.8,\quad b=0.2,\quad x=0.5,\quad
\pi=0.9,\quad \rho=0.2,\quad K=0.4,
\]
and set \(v=1\) and \(\kappa=0.5\). Then
\[
\eta_a=0.6,
\qquad
\eta_{\Hard}=0.48,
\]
so the two lower bounds in \eqref{eq:continuous-K} are
\[
m_{\Easy}\eta_a=0.156,
\qquad
\pi a\eta_a-m_{\Hard}\eta_{\Hard}=0.0768,
\]
against an upper bound of \(\pi a\eta_a=0.432\); hence \(0.156<K=0.4<0.432\). The difficult evaluator's net gain from diagnosis under hidden difficulty is \(0.032\), whereas under public difficulty no diagnosis yields the gross success gain \(m_{\Hard}\eta_{\Hard}=0.3552\). The benchmark reversal is therefore strict, though the first figure shows how thin the margin can be: with \(K\) close to \(\pi a\eta_a\) the difficult evaluator is only just willing to buy.
\end{example}

\section{Imperfect diagnostic feedback}
\label{sec:noisy}

Perfect revelation is a convenience, and an unattractive one if the mechanism depends on it. Returning to binary effort, suppose the diagnostic produces \(y\in\{h,\ell\}\) and is correct with probability \(r\in(1/2,1]\):
\[
\Prb(h\mid\theta=\High)=\Prb(\ell\mid\theta=\Low)=r.
\]
Conditional on adoption, the posterior probabilities of high ability are
\begin{align}
q_h(r)
&\equiv
\Prb(\theta=\High\mid h)
=\frac{\pi r}{\pi r+(1-\pi)(1-r)}, \label{eq:qh}\\
q_\ell(r)
&\equiv
\Prb(\theta=\High\mid \ell)
=\frac{\pi(1-r)}{\pi(1-r)+(1-\pi)r}. \label{eq:ql}
\end{align}
At any posterior \(q\) about high ability the marginal return to effort is \(b+(a-b)q\) on the difficult task and \(a-(a-b)q\) on the easy one, so as \(r\) approaches one the outcome \(h\) motivates effort on a difficult task and \(\ell\) motivates it on an easy one, as in Proposition~\ref{prop:reversal}.

If adoption is read as revealing a difficult task and only \(h\) induces effort, the difficult evaluator's success gain is
\begin{equation}
\label{eq:noisy-GH}
G_{\Hard}^{h}(r)
\equiv
\pi r a+(1-\pi)(1-r)b,
\end{equation}
the first term coming from high-ability agents who receive the favorable result and the second from false favorable results among low-ability agents. An easy evaluator who imitates obtains
\begin{equation}
\label{eq:noisy-GE}
G_{\Easy}^{h}(r)
\equiv
\pi r b+(1-\pi)(1-r)a.
\end{equation}
The two expressions behave quite differently in \(r\). One of them is bounded away from profitability for every accuracy level, since \(G_{\Easy}^{h}(r)\leq m_{\Easy}<K\) for all \(r\in[1/2,1]\); the other satisfies \(\lim_{r\to1}G_{\Hard}^{h}(r)=\pi a>K\).

\begin{theorem}
\label{thm:noisy}
Suppose Assumptions~\ref{ass:crossing}--\ref{ass:cost} hold. There exists \(\underline r<1\) such that for every \(r\in(\underline r,1]\):
\begin{enumerate}[label=(\roman*)]
 \item the unique D1 equilibrium has the easy evaluator choose no diagnostic and the difficult evaluator diagnose;
 \item after diagnosis, outcome \(h\) induces effort and outcome \(\ell\) does not;
 \item neither evaluator diagnoses when task difficulty is public;
 \item a state-independent commitment policy prescribes no diagnostic.
\end{enumerate}
\end{theorem}

The easy-type half of the refinement argument does not need continuity. Whatever the accuracy and whatever mapping from outcomes to effort the agent might use, the easy evaluator's largest possible success gain is still \(m_{\Easy}<K\), so she can never benefit from the deviation. Accuracy sufficiently close to one is used both to obtain the on-path effort pattern and to ensure \(G_{\Hard}^{h}(r)>K\) for the difficult type.

In Example~\ref{ex:numeric}, take \(r=0.95\). Then
\[
q_h(0.95)\simeq0.994,
\qquad
q_\ell(0.95)\simeq0.321,
\]
and the corresponding marginal returns to effort on the difficult task are approximately \(0.797\) and \(0.393\), on opposite sides of \(x=0.5\). Moreover
\[
G_{\Hard}^{h}(0.95)=0.685>K=0.4,
\]
so the separating equilibrium survives both a pronounced prior asymmetry and a nonzero error rate with a wide margin.

\section{When unfavorable feedback motivates}
\label{sec:symmetry}

The baseline emphasizes favorable feedback only because high ability is the modal type. Reverse that and everything reverses with it, exactly. Suppose low ability is common, \(\pi<1/2\), and easy tasks are rare. Define \(m_{\Easy}\), \(m_{\Hard}\), and \(D_\theta(\rho)\) as before and impose
\begin{equation}
\label{eq:symmetric-assumptions}
m_{\Hard}<x<m_{\Easy},
\qquad
D_0(\rho)<x,
\qquad
D_{\Low}(\rho)<x<D_{\High}(\rho),
\end{equation}
together with
\begin{equation}
\label{eq:symmetric-cost}
m_{\Hard}<K<(1-\pi)a.
\end{equation}
The first condition in \eqref{eq:symmetric-assumptions} is the counterpart of Assumption~\ref{ass:crossing} and holds because low ability is the modal type; the remaining two are the counterparts of Assumption~\ref{ass:rare} and require difficult tasks to be sufficiently common, so that the prior now resembles the difficult environment.

\begin{corollary}
\label{cor:symmetry}
Under \eqref{eq:symmetric-assumptions}--\eqref{eq:symmetric-cost}, the unique D1 equilibrium has only the easy-task evaluator purchase the diagnostic. Adoption reveals an easy task; an unfavorable result induces effort and a favorable one does not. Neither type diagnoses under public difficulty or state-independent commitment.
\end{corollary}

The labels ``favorable'' and ``unfavorable'' do not drive the result. The evaluator diagnoses the rare task on which effort is productive for the modal ability type, and a result confirming that type is then the one that identifies the agents who should work: favorable feedback on a difficult task when high ability is common, unfavorable feedback on an easy task when low ability is common.\footnote{The corollary is a relabeling rather than a new argument: exchanging the two ability labels and the two task labels maps \eqref{eq:symmetric-assumptions}--\eqref{eq:symmetric-cost} onto Assumptions~\ref{ass:crossing}--\ref{ass:cost}, with \(\pi\) replaced by \(1-\pi\) and \(\rho\) by \(1-\rho\). The model therefore has no built-in bias toward encouragement.}

\section{Welfare and empirical implications}
\label{sec:welfare}

Over-adoption is a statement about behavior, not about desirability, and the two should not be run together. This section separates them and then draws out what the model implies for data.

\subsection{Welfare}

We report welfare for the baseline binary-effort model; the continuous-effort analogue simply replaces unit effort increments with the efforts of Section~\ref{sec:continuous}. Let total surplus be the sum of the two players' payoffs, treating \(K\) as a real resource cost, so that the social value of a success is \(1+v\). Baseline success probabilities are the same across diagnostic regimes, so we again compare incremental surplus.

Under pooling without diagnosis, Assumption~\ref{ass:rare} gives no effort and hence zero incremental surplus. In the hidden-difficulty D1 equilibrium only high-ability agents work, and only on the difficult task, where incremental surplus is
\begin{equation}
\label{eq:Whidden}
W_{\Hard}^{\mathrm{hidden}}
=\pi\big[(1+v)a-c\big]-K.
\end{equation}
This is positive, and comfortably so: \(K<\pi a\) and \(va>c\) together give
\[
W_{\Hard}^{\mathrm{hidden}}
>
\pi\big[(1+v)a-c\big]-\pi a
=\pi(va-c)>0.
\]

If difficulty is public, everyone works on the difficult task and nobody works on the easy one, so conditional on a difficult task incremental surplus is
\begin{equation}
\label{eq:Wpublic}
W_{\Hard}^{\mathrm{public}}
=(1+v)m_{\Hard}-c,
\end{equation}
and the public-minus-hidden difference is
\begin{equation}
\label{eq:welfare-diff}
W_{\Hard}^{\mathrm{public}}-W_{\Hard}^{\mathrm{hidden}}
=(1-\pi)\big[(1+v)b-c\big]+K.
\end{equation}

\begin{proposition}
\label{prop:welfare}
Under Assumptions~\ref{ass:crossing}--\ref{ass:cost}:
\begin{enumerate}[label=(\roman*)]
 \item the hidden-difficulty D1 equilibrium strictly improves total surplus relative to the no-diagnostic pooling equilibrium;
 \item the ranking between hidden and public difficulty is ambiguous, and is settled by the sign of \eqref{eq:welfare-diff}.
\end{enumerate}
\end{proposition}

Diagnostic over-adoption thus carries no automatic welfare implication in either direction. Hidden difficulty improves on uninformative pooling by generating productive effort among high-ability agents. Public difficulty saves the diagnostic fee and additionally puts the low-ability agents to work on the difficult task. Which dominates depends on whether the social value of that extra effort, \((1+v)b\), covers its cost \(c\)---and on how large the fee is.\footnote{Both signs occur inside Assumptions~\ref{ass:crossing}--\ref{ass:cost}, so the ambiguity is real rather than an artifact of leaving parameters unrestricted. Since \(c=vx\) and \(b<x\), the bracket in \eqref{eq:welfare-diff} is negative whenever \(v(x-b)>b\); when in addition \((1-\pi)\big[v(x-b)-b\big]>K\), hidden difficulty is the better regime despite the fee. A high-stakes task with a large \(v\), on which mismatched agents gain very little from effort, is the case in which concealing difficulty is not merely privately but socially preferable.}

\subsection{Empirical content}

The model's predictions are about interactions---between adoption, difficulty, and the result---rather than about any of the three separately, which is what makes them testable.

\begin{enumerate}[label=(\roman*)]
 \item \textit{Adoption and difficulty.} Where high ability is common, assessments should be disproportionately adopted for unusually difficult assignments, even after conditioning on the direct cost of assessing.
 \item \textit{Adoption changes the meaning of a result.} A favorable result should raise post-assessment effort when the adoption decision was itself informative about a difficult task. On a known easy task, effort should instead be higher after an unfavorable than after a favorable result.
 \item \textit{Transparency can reduce assessment.} Making task difficulty public can reduce adoption, including on difficult tasks, because unresolved ability uncertainty was motivating a broader set of agents.
 \item \textit{Population composition reverses feedback.} As the modal ability type changes, both the task on which assessments are adopted and the sign of the motivating result should reverse, as in Corollary~\ref{cor:symmetry}.
 \item \textit{Standardization removes the adoption signal.} A central requirement imposing the same diagnostic policy on all tasks should produce less assessment than decentralized evaluators who privately observe difficulty.
\end{enumerate}

Two caveats govern all five. They concern assessments administered \emph{before} productive effort, not feedback delivered once effort is sunk; and they require an independent measure of task difficulty, since the whole content of the mechanism is that the same result means different things depending on what adoption revealed. A useful setting would combine that measure with variation in adoption across informed evaluators and an observable measure of subsequent preparation. The diagnostic result should then be interacted with adoption and with difficulty jointly: what the model predicts is a change in the \emph{sign} of the result--effort relationship, not merely a change in its magnitude.\footnote{A competing explanation is that evaluators simply value information more on difficult tasks. Distinguishing the mechanisms therefore requires variation in task transparency, together with controls for the diagnostic's direct informational value. The model's sharper prediction is that making difficulty public eliminates adoption in the stated parameter region.}

\section{Conclusion}
\label{sec:conclusion}

When an informed evaluator decides whether to diagnose an agent, the decision itself can disclose something about the task. In the model studied here that disclosure is what makes the diagnostic worth buying: neither evaluator type buys it when difficulty is public, and a policy that cannot condition on difficulty prescribes not buying it either.

What generates this is a crossing return to effort, under which favorable feedback motivates on a difficult task while unfavorable feedback can motivate on an easy one. When high ability is common and difficult tasks are rare, only the difficult evaluator diagnoses; D1 selects that outcome because an unexpected diagnostic can never benefit the easy evaluator and can benefit the difficult one, and no randomization is available to either. The conclusion survives continuous effort and sufficiently accurate diagnostics, and the core mechanism does not require symmetric returns to effort. Reverse the modal ability type and both the diagnosed task and the motivating result reverse with it.

Two things are worth keeping separate in reading this. What is established is a comparison of behavior across informational regimes: adoption is positive under hidden difficulty and zero under both benchmarks. What is not established is a welfare ranking---the comparison with transparency goes either way, depending on the social value and cost of the additional low-ability effort and on the diagnostic fee. Transparency about the task and information about the agent, in other words, need not be complements. Concealing difficulty can generate more information about the person because the act of seeking it becomes informative about the task.

\appendix

\section{Proofs}
\label{app:proofs}

\begin{proof}[Proof of Lemma~\ref{lem:foundation}]
The argument is that \(\Delta\) is a single-peaked function of the index, so equal distances from the peak give equal increments.

Symmetry of \(F\) gives \(F(-y)=1-F(y)\), and hence
\[
\Delta(-\gamma-q)
=F(-q)-F(-\gamma-q)
=F(q+\gamma)-F(q)
=\Delta(q).
\]
Since \(q\mapsto-\gamma-q\) is reflection about \(-\gamma/2\), the function \(\Delta\) is symmetric about that point. For the peak, differentiate:
\[
\Delta'(q)=f(q+\gamma)-f(q).
\]
If \(q>-\gamma/2\) then \(q+\gamma>-q\) and \(q+\gamma>0\), so \(|q+\gamma|>|q|\); symmetry and strict unimodality of \(f\) then give \(f(q+\gamma)<f(q)\), so \(\Delta'(q)<0\). The reverse holds for \(q<-\gamma/2\). Hence \(\Delta\) is uniquely maximized at \(q=-\gamma/2\) and strictly decreasing in \(|q+\gamma/2|\).

It remains to locate the four indices. With the stated thresholds, \(q=\alpha_\theta-T_t\) equals \(-\gamma/2\) for \((\Easy,\Low)\) and for \((\Hard,\High)\), which gives the common value \(a\). Writing \(g=\alpha_{\High}-\alpha_{\Low}>0\), the two mismatched indices are \(g-\gamma/2\) and \(-g-\gamma/2\), at equal distance \(g\) from the peak. They therefore generate the same increment, and it is strictly below \(a\).
\end{proof}

\begin{proof}[Proof of Proposition~\ref{prop:reversal}]
Each claim is a comparison of one number with \(x\). On a known easy task a revealed low-ability agent faces marginal return \(a>x\) and a revealed high-ability agent faces \(b<x\); on a known difficult task the two are \(b<x\) and \(a>x\) respectively. Without a diagnostic the returns are \(m_{\Easy}<x\) and \(m_{\Hard}>x\). All four inequalities are Assumption~\ref{ass:crossing}.
\end{proof}

\begin{proof}[Proof of Proposition~\ref{prop:cheap-talk}]
Suppose there were a fully revealing equilibrium. A message interpreted as \(\Hard\) induces effort, since \(m_{\Hard}>x\), and a message interpreted as \(\Easy\) induces none, since \(m_{\Easy}<x\). The easy evaluator, assigned the latter, can send the former instead at no cost and gain \(m_{\Easy}>0\), because effort raises her success probability whatever the task. So full revelation cannot be sustained. In a babbling equilibrium the message leaves the task posterior at \(\rho\), and \(D_0(\rho)<x\) gives no effort.
\end{proof}

\begin{proof}[Proof of Proposition~\ref{prop:separating}]
Under \eqref{eq:sep-policy}, Bayes' rule gives \(\mu_0=0\) and \(\mu_1=1\). After \(z=0\) the expected marginal return is \(m_{\Easy}<x\), so \(e=0\); after \(z=1\) it is \(a>x\) for \(\theta=\High\) and \(b<x\) for \(\theta=\Low\), so the stated effort strategy is sequentially rational.

For the evaluators, compare each type's payoff with what she would obtain by deviating to \(z=0\) and inducing no effort. The difficult type gains \(\pi a-K>0\) by Assumption~\ref{ass:cost}. The easy type, imitating \(z=1\), gains \(\pi b-K<m_{\Easy}-K<0\), the first inequality because \(m_{\Easy}=(1-\pi)a+\pi b>\pi b\). Both types therefore follow \eqref{eq:sep-policy}.
\end{proof}

\begin{proof}[Proof of Proposition~\ref{prop:pbe-set}]
Part~(i) is Proposition~\ref{prop:separating}. For part~(ii), pooling on \(z=0\) gives \(\mu_0=\rho\) by Bayes' rule and hence \(e=0\) by Assumption~\ref{ass:rare}. Assign \(\mu_1=0\) after an off-path diagnostic, so that only an unfavorable result induces effort. The deviation gains are then
\[
(1-\pi)a-K
\quad\text{and}\quad
(1-\pi)b-K
\]
for the easy and difficult types respectively, and both are negative because \(K>m_{\Easy}=(1-\pi)a+\pi b>(1-\pi)a>(1-\pi)b\). Pooling on no diagnosis is therefore a PBE.

For part~(iii), take the two remaining pure profiles in turn. If both types diagnose, the posterior stays at \(\rho\) and Assumption~\ref{ass:rare} implies that only low ability works, so the easy evaluator's incremental payoff is \((1-\pi)a-K<0\). Deviating to no diagnosis saves \(K\) and cannot make her worse off, since any effort induced after that deviation only raises her success probability. So pooling on diagnosis fails. If only the easy evaluator diagnoses, adoption is read as \(t=\Easy\) and again only low ability works, leaving her with \((1-\pi)a-K<0\) against a weakly positive payoff from not diagnosing. The reverse separating profile fails too. These four profiles exhaust the pure diagnostic strategies of the two types.
\end{proof}

\begin{proof}[Proof of Proposition~\ref{prop:no-mixing}]
Take the two types in turn; neither can be made indifferent, though for different reasons.

\emph{The easy type.} Because baseline success does not depend on \(z\), the difference between her payoffs from \(z=1\) and \(z=0\) is the difference in the success gain created by effort, less \(K\). Fix any beliefs and any continuation the agent might play after \(z=1\), including a randomized one, and let \(\lambda_{\Low},\lambda_{\High}\in[0,1]\) be the probabilities with which the two revealed ability types work. Her gain from effort is then
\[
(1-\pi)\lambda_{\Low}a+\pi\lambda_{\High}b
\leq
(1-\pi)a+\pi b
=m_{\Easy},
\]
attained when both types work with certainty. Her gain from effort after \(z=0\) is nonnegative. Hence
\[
u_P(z=1)-u_P(z=0)\leq m_{\Easy}-K<0
\]
by Assumption~\ref{ass:cost}. No diagnostic strictly dominates diagnosing for her, at every belief and against every continuation, so she assigns \(z=1\) probability zero in every PBE.

\emph{The difficult type.} Suppose she diagnoses with probability \(\sigma\in(0,1)\). Both decisions are then on path, and by the previous step she is the only type supplying a diagnostic, so \(\mu_1=1\) and
\[
\mu_0=\frac{\rho(1-\sigma)}{\rho(1-\sigma)+1-\rho}\leq\rho.
\]
Since \(m_{\Hard}>m_{\Easy}\), the function \(D_0\) is strictly increasing, so \(D_0(\mu_0)\leq D_0(\rho)<x\) by Assumption~\ref{ass:rare} and no effort follows \(z=0\): her gain there is zero. After \(z=1\) the agent knows both the task and her ability, so high ability works and low ability does not, and her gain is \(\pi a-K>0\). She therefore strictly prefers \(z=1\), contradicting the indifference that \(\sigma\in(0,1)\) requires.

Both types thus play pure strategies, so every PBE is one of the four profiles enumerated in Proposition~\ref{prop:pbe-set}, two of which are ruled out by part~(iii) of that proposition.
\end{proof}

\begin{proof}[Proof of Theorem~\ref{thm:d1}]
By Proposition~\ref{prop:no-mixing} the equilibrium set is exactly \{separating, pooling on no diagnosis\}, so it suffices to show that pooling fails D1 and that separation does not.

At the pooling PBE consider the off-path action \(z=1\). By the first step of the proof of Proposition~\ref{prop:no-mixing}, under any continuation response the easy evaluator's gross success gain is at most \(m_{\Easy}\), and \(m_{\Easy}-K<0\); there is therefore no response, randomized or not, under which she weakly benefits from the deviation. For the difficult evaluator the continuation belief \(\mu_1=1\) makes only a favorable result induce effort, for the strictly positive gain \(\pi a-K\). The set of receiver responses that would make the deviation profitable is thus empty for the easy type and nonempty for the difficult one, so D1 assigns probability zero to the easy type after \(z=1\). Given the resulting belief \(\mu_1=1\), the difficult type deviates profitably, and pooling fails.

Separation satisfies D1 trivially, since both actions are on path and Bayes' rule pins both beliefs. It is therefore the unique PBE satisfying D1.
\end{proof}

\begin{proof}[Proof of Proposition~\ref{prop:region}]
The cost interval is nonempty exactly when
\[
m_{\Easy}=(1-\pi)a+\pi b<\pi a,
\]
and rearranging gives \(a<\pi(2a-b)\), which is \eqref{eq:pi-bound}. The endpoints move as claimed because the lower one has derivative \(-(a-b)<0\) in \(\pi\) and the upper one has derivative \(a>0\). For part~(iii), the three inequalities of Assumption~\ref{ass:rare} are respectively
\[
\rho<\frac{x-m_{\Easy}}{m_{\Hard}-m_{\Easy}},
\qquad
\rho<\frac{x-b}{a-b},
\qquad
\rho<\frac{a-x}{a-b},
\]
whose conjunction is \(\rho<\bar\rho\).
\end{proof}

\begin{proof}[Proof of Proposition~\ref{prop:public}]
On a public easy task, no diagnosis induces no effort and diagnosis induces effort only from low ability, so the evaluator's gain from diagnosing is \((1-\pi)a-K<m_{\Easy}-K<0\).

On a public difficult task, no diagnosis induces both ability types to work, for incremental success \(m_{\Hard}\); diagnosis induces only high ability to work, for \(\pi a-K\). By \eqref{eq:public-hard-diff} the former exceeds the latter by \((1-\pi)b+K>0\).
\end{proof}

\begin{proof}[Proof of Proposition~\ref{prop:commitment}]
With no diagnostic the task posterior is \(\rho\), so Assumption~\ref{ass:rare} implies no effort and a gain of zero. With a state-independent diagnostic, \(D_{\Low}(\rho)>x>D_{\High}(\rho)\), so only low ability works and the ex ante gain is \eqref{eq:commit-gain}, which is strictly negative by the inequality displayed after it. A task-independent random diagnostic delivers that negative gain in proportion to the probability with which it is used, so zero is optimal.
\end{proof}

\begin{proof}[Proof of Corollary~\ref{cor:overadoption}]
In the equilibrium selected by Theorem~\ref{thm:d1} the diagnostic is bought exactly on the difficult task, which occurs with probability \(\rho>0\). Propositions~\ref{prop:public} and~\ref{prop:commitment} give adoption probability zero in the two benchmarks.
\end{proof}

\begin{proof}[Proof of Theorem~\ref{thm:continuous}]
Given perceived marginal productivity \(q\), the agent solves
\[
\max_{e\in[0,\bar e]}
\left\{v(q-x)e-\frac{\kappa}{2}e^2\right\},
\]
and since the upper bound is slack by assumption the unique solution is \(\eta(q)\) in \eqref{eq:continuous-br}. Note for later use that perceived productivity always lies in \([b,a]\), so no continuation can elicit effort above \(\eta_a\).

\emph{Part~(i), hidden difficulty.} Under the separating policy \eqref{eq:sep-policy}, Bayes' rule gives \(\mu_0=0\) and \(\mu_1=1\). No diagnosis then induces \(\eta(m_{\Easy})=0\); after diagnosis high ability supplies \(\eta(a)=\eta_a\) and low ability supplies \(\eta(b)=0\). The difficult evaluator's gain is \(\pi a\eta_a-K>0\) by \eqref{eq:continuous-K}, and an easy evaluator who imitates gains \(\pi b\eta_a-K<m_{\Easy}\eta_a-K<0\). The separating profile is a PBE.

Pooling on no diagnosis can again be supported by attributing an unexpected diagnostic to the easy type: the prior induces zero effort because \(D_0(\rho)<x\), and following the off-path diagnostic low ability supplies \(\eta_a\) while high ability supplies zero, giving deviation gains \((1-\pi)a\eta_a-K<0\) and \((1-\pi)b\eta_a-K<0\). Pooling on diagnosis is impossible, since under the prior only low ability supplies positive effort and that effort is at most \(\eta_a\), so the easy evaluator's gross gain is at most \((1-\pi)a\eta_a<m_{\Easy}\eta_a<K\); the reverse separating profile fails by the same bound. Randomization is unavailable for the reasons given in Proposition~\ref{prop:no-mixing}: the easy type's gross gain from diagnosing is at most \(m_{\Easy}\eta_a<K\) under every continuation, and if the difficult type randomized then \(\mu_0\leq\rho\) would give \(\eta(D_0(\mu_0))=0\) and a gain of zero against \(\pi a\eta_a-K>0\).

For D1, any sequentially rational effort after diagnosis is at most \(\eta_a\), so the easy evaluator's gross gain is at most
\[
\big[(1-\pi)a+\pi b\big]\eta_a
=m_{\Easy}\eta_a<K,
\]
and she cannot gain under any continuation response, while the difficult type gains when the deviation is attributed to her. D1 therefore eliminates pooling, which establishes part~(i).

\emph{Part~(ii), public difficulty.} On a public easy task, diagnosis induces effort \(\eta_a\) only from low ability, for a gain of \((1-\pi)a\eta_a-K<m_{\Easy}\eta_a-K<0\). On a public difficult task, no diagnosis induces \(\eta_{\Hard}\) from everyone and gross success \(m_{\Hard}\eta_{\Hard}\), while diagnosis induces \(\eta_a\) from high ability alone and yields \(\pi a\eta_a-K\). The second lower bound in \eqref{eq:continuous-K} is exactly the statement that the former is strictly larger.

\emph{Part~(iii), commitment.} No diagnosis leaves the posterior at \(\rho\) and induces zero effort. A state-independent diagnostic induces effort only after an unfavorable result, at level \(\eta(D_{\Low}(\rho))\), for an ex ante gain of \((1-\pi)D_{\Low}(\rho)\eta(D_{\Low}(\rho))-K\). Since \(D_{\Low}(\rho)\leq a\) and \(\eta\) is nondecreasing, this is at most \((1-\pi)a\eta_a-K<m_{\Easy}\eta_a-K<0\).

\emph{Nonemptiness.} The upper endpoint of \eqref{eq:continuous-K} exceeds the first lower component exactly when \(m_{\Easy}<\pi a\), and exceeds the second by \(m_{\Hard}\eta_{\Hard}>0\), the latter because \(m_{\Hard}>x\). The interval is therefore nonempty precisely under \eqref{eq:pi-bound}.
\end{proof}

\begin{proof}[Proof of Theorem~\ref{thm:noisy}]
The proof separates a continuity argument, which delivers the on-path effort pattern, from a global argument, which delivers the refinement.

\emph{Continuity.} At \(r=1\) every effort comparison is strict, by Assumptions~\ref{ass:crossing} and~\ref{ass:rare}, and the posteriors \eqref{eq:qh}--\eqref{eq:ql} and all induced marginal returns are continuous in \(r\). Hence for \(r\) sufficiently close to one, adoption interpreted as \(\Hard\) makes \(h\) induce effort and \(\ell\) induce none. The difficult evaluator's adoption gain is then \(G_{\Hard}^{h}(r)-K\), which converges to \(\pi a-K>0\), while the easy evaluator's imitation gain is \(G_{\Easy}^{h}(r)-K<0\) because \(G_{\Easy}^{h}(r)\leq m_{\Easy}<K\). The separating PBE persists. Pooling on no diagnosis also persists, since the deviation gains computed in the proof of Proposition~\ref{prop:pbe-set} are strictly negative at \(r=1\) and continuity preserves them nearby; the other two pure profiles remain impossible for the same reason, and randomization remains unavailable by the argument of Proposition~\ref{prop:no-mixing}.

\emph{Global step.} The bound on the easy evaluator does not use continuity at all. Whatever the accuracy \(r\) and whatever mapping from outcomes to effort the agent uses, the low- and high-ability contributions to her gross success gain are bounded respectively by \((1-\pi)a\) and \(\pi b\). Their sum is therefore at most \(m_{\Easy}<K\), so she can never benefit from the deviation. The difficult type benefits whenever \(G_{\Hard}^{h}(r)>K\), which holds near \(r=1\). D1 again attributes an unexpected diagnostic to her and eliminates pooling.

\emph{Benchmarks.} With difficulty public, diagnosis cannot improve the easy evaluator's success by more than \(m_{\Easy}<K\); on the difficult task no diagnosis already induces both types to work, and a diagnostic can induce at most the same effort profile while costing \(K\), so neither type diagnoses. For commitment, the state-independent diagnostic has strictly negative value at \(r=1\) by Proposition~\ref{prop:commitment}, and both that value and the continuation effort inequalities are continuous in \(r\), so no diagnostic remains optimal near one. Taking the largest of the finitely many accuracy bounds obtained above yields a common \(\underline r<1\).
\end{proof}

\begin{proof}[Proof of Corollary~\ref{cor:symmetry}]
Relabel ability \(\Low\) as \(\High\) and \(\High\) as \(\Low\), and the easy task as difficult and the difficult task as easy. The crossing matrix \eqref{eq:increments} is invariant under this double exchange, while \(\pi\) becomes \(1-\pi>1/2\) and \(\rho\) becomes \(1-\rho\). Under the relabeling, \(m_{\Easy}\) and \(m_{\Hard}\) exchange places, as do \(D_{\Low}(\rho)\) and \(D_{\High}(\rho)\), while \(D_0(\rho)\) is unchanged; conditions \eqref{eq:symmetric-assumptions}--\eqref{eq:symmetric-cost} therefore become Assumptions~\ref{ass:crossing}--\ref{ass:cost} exactly. The conclusion follows from Theorem~\ref{thm:d1} and Propositions~\ref{prop:public} and~\ref{prop:commitment}, read in the relabeled model.
\end{proof}

\begin{proof}[Proof of Proposition~\ref{prop:welfare}]
Conditional on a difficult task, the hidden-difficulty equilibrium creates incremental success \(\pi a\), which is worth \(\pi a\) to the evaluator and \(v\pi a\) to the agent, against expected effort cost \(\pi c\) and diagnostic cost \(K\); this is \eqref{eq:Whidden}, and the display following that equation shows it is strictly positive. Multiplying by \(\rho>0\) gives a strict ex ante improvement over pooling without diagnosis, which generates no incremental surplus at all. That is part~(i).

For part~(ii), under public difficulty everyone works on the difficult task, so incremental success is \(m_{\Hard}\) and effort cost is \(c\), giving \eqref{eq:Wpublic}. Subtracting,
\[
(1+v)(m_{\Hard}-\pi a)-c+\pi c+K
=(1-\pi)\big[(1+v)b-c\big]+K,
\]
using \(m_{\Hard}-\pi a=(1-\pi)b\). The bracket is positive for small \(v\) and negative for large \(v\), since \(c=vx\) with \(x>b\), so the whole expression takes either sign within Assumptions~\ref{ass:crossing}--\ref{ass:cost}.
\end{proof}

\end{document}